\documentclass[a4paper,11pt]{article}
\pdfoutput=1 

\usepackage{jinstpub}

\usepackage{amsmath}

\usepackage{threeparttable}
\graphicspath{{used_figures/}}
\title{\boldmath An analytical model for electronic noise in a cryogenic bolometer detector readout circuit}

\author[a,b]{V. Vatsa,}
\author[c,1]{\note{Present address: Fermi National Accelerator Laboratory, Illinois-60510, USA}A. Reza,}
\author[a,b,2]{\note{Present address: Los Alamos National Laboratory, New Mexico-87545, USA}A. Mazumdar,}
\author[c]{M. S. Pose,}
\author[c]{S. Mallikarjunachary,}
\author[c,3]{\note{Corresponding author}V.~Nanal,}
\author[d]{R. G. Pillay,}
\author[e,4]{\note{Present address: Dept. of Physics, IISER Pune, Pune-411008, India}S. Ramakrishnan}
\author[a,f]{and A. Shrivastava}

\affiliation[a]{Homi Bhabha National Institute,\\Mumbai-400094, India}
\affiliation[b]{INO, Tata Institute of Fundamental Research,\\Mumbai-400005, India}
\affiliation[c]{DNAP, Tata Institute of Fundamental Research,\\Mumbai-400005, India}
\affiliation[d]{Department of Physics, Indian Institute of Technology, \\ Ropar-140001, India
}
\affiliation[e]{DCMP\&MS, Tata Institute of Fundamental Research,\\Mumbai-400005, India}
\affiliation[f]{Nuclear Physics Division, Bhabha Atomic Research Centre, \\Mumbai 400085, India}

\emailAdd{nanal@tifr.res.in}

\abstract{This paper presents an analytical model to quantify the measured noise in a cryogenic bolometer readout circuit.  The model includes the contributions from the bias resistors and sensor resistors, voltage and current noise of amplifier, and cable capacitance. The model parameters are empirically estimated using frequency domain analysis of the measured noise data of indigenously developed Neutron Transmutation Doped (NTD) Ge sensors. The model is  shown to describe noise data for NTD Ge sensors 
over a wide range of resistances corresponding to temperatures in the range 20~-~70~mK.
Relative contributions of different components are discussed and it is shown that the contribution to the overall noise from the differential amplifier at 300~K is the dominant source. It is observed that the amplifier flicker noise is significantly lower than that specified in the amplifier datasheet. The present study also indicates that  a desirable value of resistance of NTD sensor (\textit{$R_{NTD}$}) from noise considerations is $\lesssim$ 1~G$\Omega$  at  $\sim$~20~mK. 
}

\keywords{Cryogenic detectors, Double-beta decay detectors, Front-end electronics for detector readout}

\begin{document}
\maketitle
\flushbottom

\section{Introduction}
A cryogenic bolometer detector 
offers an excellent energy resolution and high sensitivity, which makes it an ideal choice for several rare decay experiments such as neutrinoless double beta decay (NDBD) and dark matter searches~\cite{ndm_dark}. Ideally the performance of a bolometer detector is expected to be superior due to its good intrinsic resolution.

However, in practice, resolution is often limited by the noise contributed from other sources~\cite{low_temp_det}. Understanding these external noise contributions and mitigating the same is of paramount importance.
Some of the major contributions arise from vibrations (e.g. pulse tube cooler ~\cite{PT1,PT2}), the Johnson noise in the sensor~\cite{Cuore1} and other electronic components in the bolometer readout circuit. Minimization of the electrical noise in the readout circuit has attracted much attention and the efforts have been concentrated towards the design of low noise preamplifier ~\cite{cryo_amp} and preamplifiers operating at low temperature~\cite{ashif_LTD}. The  effect of external noise pickups on the 
performance of the bolometer is already reported in Ref.~\cite{garai_noise}. 
A typical readout circuit consists of a  Neutron Transmutation Doped (NTD) Ge sensor at mK temperature, a current source for sensor biasing (generated by applying a constant voltage through a high bias resistor) and a front-end preamplifier. 
In order to understand the effect of noise from each of these sources and to predict the noise for a given readout circuit, a noise analysis of the bolometer readout circuit is required. An analytical noise model can also provide key information for the design and optimization of low noise readout circuitry for the bolometer.  

A tin cryogenic bolometer detector (\textit{TIN.TIN}-The INdia-based TIN detector)~\cite{nanal_epj} is being developed to search for NDBD process in $^{124}$Sn in the upcoming India-based Neutrino Observatory (INO). At Tata Institute of Fundamental Research, Mumbai, a prototype bolometer test setup with a sapphire absorber and indigeneously fabricated NTD Ge sensor~\cite{garai_bolometer} has been developed for initial testing in a cryogen-free dilution refrigerator CFDR-1200 (Leiden Cryogenics)~\cite{v_singh}. 

In this article, we present an analytical model for the estimation of electronic noise in a bolometer readout circuit. The model parameters are empirically determined from the fits to noise data of a family of NTD Ge sensors in a limited temperature range $T\sim$ 40~-~70 mK.  
Since NTD Ge sensors have a limited operational range due to a very steep temperature dependence   of R, noise measurements have also been carried out with standard surface mount device (SMD) 
resistors, where R values could be chosen flexibly to cover a higher temperature range.
The simulated noise spectra with optimized parameters are compared with  measured noise of NTD Ge sensors and SMD resistors. 
Further, it is shown that the amplifier  noise is the predominant source in most cases. The analysis also shows that the $R_{NTD}$ should be constrained below 1~G$\Omega$ to keep the thermal noise of the sensor as well as overall noise within acceptable limits.

\section{Analytical model for the bolometer noise}
\label{section2}
\begin{figure}
\begin{center}
	\includegraphics[scale = 0.7]{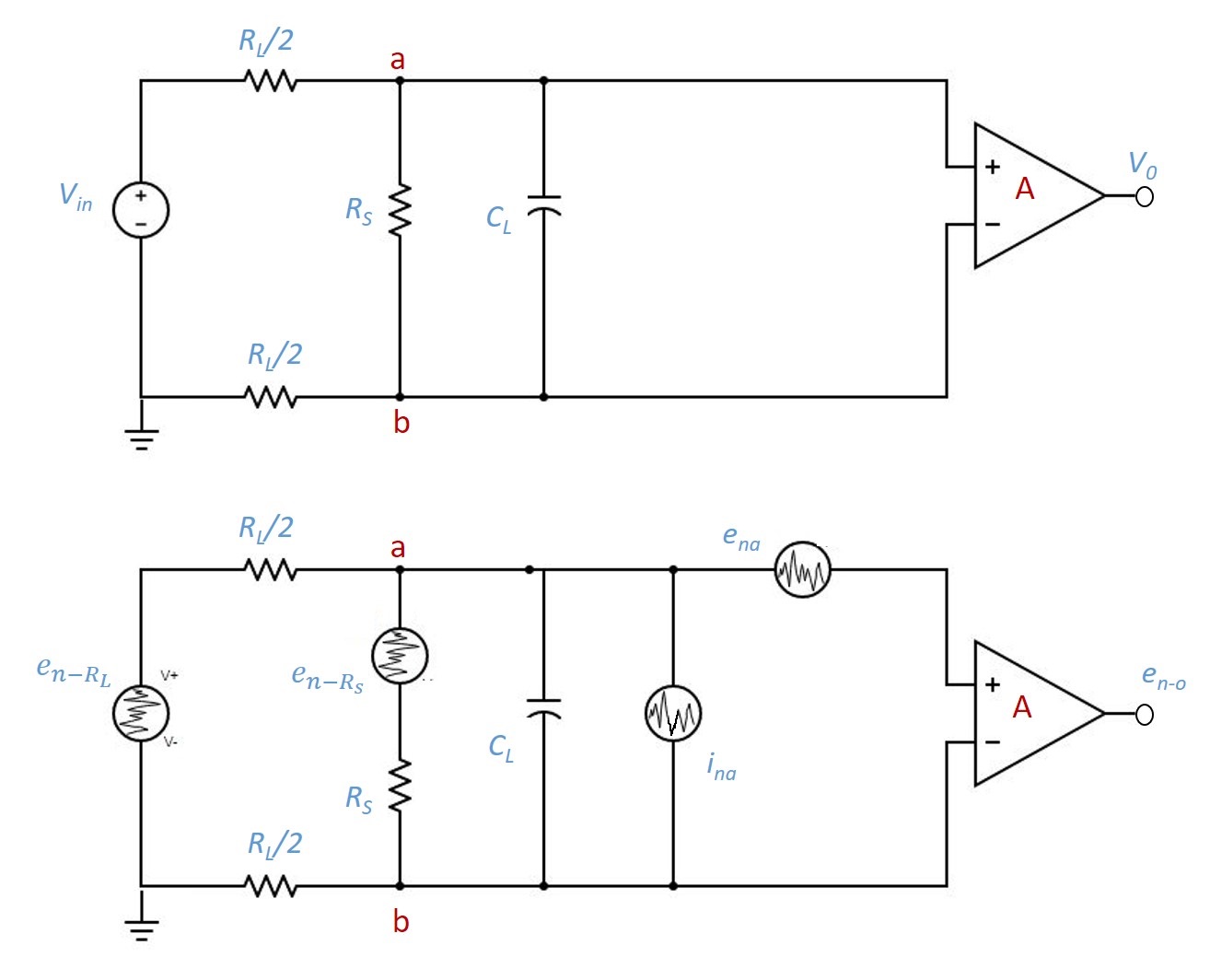}
	\caption{A bolometer readout circuit (top panel) and its equivalent noise model (bottom panel)}
	\label{fig:ckt}
\end{center}
\end{figure}

The bolometer readout circuit and its equivalent noise model are shown in Figure~\ref{fig:ckt}. A pseudo-constant current source, consisting of a voltage source $\textit{V$_{in}$}$ in series with a high bias resistor \textit{R$_L$} is used for sensor biasing. The voltage across the sensor is amplified using a low noise differential amplifier Femto DLPVA-100-F~\cite{femto}.  
This amplifier, referred to as FEMTO hereafter, is characterised by low input voltage noise (6.9~nV/$\sqrt{Hz}$), low input current noise (1.6~fA/$\sqrt{Hz}$), variable gain (20-80 dB) and high input impedance (1 T$\Omega$).  It should be mentioned that although the current biasing scheme employs a unipolar power supply,  the high CMRR (Common Mode Rejection Ratio)  of 120~dB of   FEMTO   amplifier ensures that the common mode voltage is suppressed, and does not saturate its  output. In principle, the unipolar supply can be replaced by a symmetric bipolar supply to improve the performance.

As is well known, the power density of noise is expressed as \textit{$e_{rms}^2$} per unit frequency (V$^2$/Hz). In the following, $e_k$ represents the noise voltage contribution of the respective component,  $\sqrt{(e_{k,rms}^2)B}$ in units of Volts, where \textit{B} is the bandwidth in Hz. The Johnson thermal noise of the bias resistor $\textit{$R_L$}$ and the sensor resistor $\textit{$R_S$}$ are defined as:
\begin{equation}
e_{ n-R_L} = \sqrt{4kT_{L}BR_L} 
\end{equation}
\begin{equation}
e_{n-R_S} = \sqrt{4kT\textsubscript{S}BR\textsubscript{S}}
\end{equation}
where, \textit{$T_L$} and \textit{$T_S$} are temperatures (in Kelvin) of the bias and the  sensor resistors, respectively, 
and \textit{k} is Boltzman constant. 
It may be noted that  the noise of the sensor bias voltage \textit{V$_{in}$} i.e. of NI-DAQ card (PXI-6281)~\cite{ni_daq}, specified to be $\sim$ 60 nV/$\sqrt{ \rm Hz}$   is negligible as compared to the thermal noise of bias resistor $\sim$ 
57.5 $\mu$V/$\sqrt{ \rm Hz}$  for $\textit{$R_L$}$ = 20~G$\Omega$ at room temperature and hence is not taken into consideration in the present work. 
Further, during the measurement and analysis, the acquisition bandwidth of FEMTO was kept at 1 kHz.
The resolution bandwidth (B) is taken to be 0.1 Hz for consistency with the measured data, as explained in the later section. 
The contribution of noise at the input of the amplifier can be divided into four parts and each part can be calculated separately. \\

(a) Noise contribution \textit{($e_L$)}  due to the bias resistor \textit{$R_L$}:
\begin{equation} \label{eq1}
\begin{split}
e_{L} & = e_{ n-R_L}\frac{R\textsubscript{S}||(1/j\omega C\textsubscript{L})}{R\textsubscript{L} + [R\textsubscript{S} || (1/j\omega C\textsubscript{L})]} \\
& = \sqrt{4kT\textsubscript{L}BR\textsubscript{L}} \frac{R\textsubscript{S}}{R\textsubscript{S} + R\textsubscript{L} + j\omega R\textsubscript{S}R\textsubscript{L}C\textsubscript{L}}
\end{split} 
\end{equation}

\begin{equation}
\begin{split}
\lvert e\textsubscript{L} \rvert & = \sqrt{4kT\textsubscript{L}BR\textsubscript{L}} \frac{R\textsubscript{S}}{\sqrt{(R\textsubscript{S}+R\textsubscript{L})\textsuperscript{2} + (\omega R\textsubscript{S}R\textsubscript{L}C\textsubscript{L})\textsuperscript{2}}} \\
& = \sqrt{4kT\textsubscript{L}BR\textsubscript{L}}\frac{1}{R_{L}}\frac{R\textsubscript{eq}}{\sqrt{1 + (\omega R\textsubscript{eq}C\textsubscript{L})\textsuperscript{2}}}
\end{split}
\end{equation}
where \textit{$R_{eq}$} is \textit{$R_{S}$} $\mid$$\mid$ \textit{$R_{L}$}. The Eq. 2.4 depicts the behaviour of  a low pass filter (LPF) with cutoff frequency $f_L$ = 1/2$\pi R_{eq}$C$_{L}$. It can be seen from Eq. 2.4 that at $\omega$ = 0, only a small fraction (R\textsubscript{S}/(R\textsubscript{S}+R\textsubscript{L})) of thermal noise from \textit{R\textsubscript{L}} reaches the input of the amplifier. However, for \textit{R\textsubscript{L}} = 20~G$\Omega$ at 300~K ($e_{n-R_L}$ = 18.2~$\mu$V) and $R_S$~=~500~M$\Omega$ (typical $R_{NTD}$ at  $T_S\sim$~35~mK), the contribution of thermal noise from $R_L$ at the input of the amplifier can be significantly larger (~$\sim$0.44$~\mu$V). For a finite cable capacitance \textit{C$_L$}, the noise contribution $e_L$  reduces with increasing frequency.\\

(b) Noise contribution \textit{($e_S$)} due to the sensor resistance $R_S$:
\begin{equation} \label{eq1}
\begin{split}
e_S & = e_{n-R_S}\frac{R\textsubscript{L}||(1/j\omega C\textsubscript{L})}{R\textsubscript{S} + [R\textsubscript{L} || (1/j\omega C\textsubscript{L})]} \\
& = \sqrt{4kT\textsubscript{S}BR\textsubscript{S}} \frac{R\textsubscript{L}}{R\textsubscript{S} + R\textsubscript{L} + j\omega R\textsubscript{S}R\textsubscript{L}C\textsubscript{L}}
\end{split}
\end{equation}

\begin{equation}
\begin{split}
\lvert e\textsubscript{S} \rvert & = \sqrt{4kT\textsubscript{S}BR\textsubscript{S}} \frac{R\textsubscript{L}}{\sqrt{(R\textsubscript{S}+R\textsubscript{L})\textsuperscript{2} + (\omega R\textsubscript{S}R\textsubscript{L}C\textsubscript{L})\textsuperscript{2}}} \\
& = \sqrt{4kT\textsubscript{S}BR\textsubscript{S}}\frac{1}{R_{S}}\frac{R\textsubscript{eq}}{\sqrt{1 + (\omega R\textsubscript{eq}C\textsubscript{L})\textsuperscript{2}}}
\end{split}
\end{equation}
The Eq. 2.6 also depicts the behaviour of a LPF. 
Almost the entire fraction  of thermal noise from \textit{$R_S$, ($R_L$/($R_S$+$R_L$)~$\sim 1$)} appears at the input of the amplifier. 
In this case also, the  $e_S$ reduces with increasing frequency for a finite cable capacitance $C_L$.\\

(c) Noise contribution \textit{($e_{ni}$)} due to amplifier input current noise density $i_{na}$:
\begin{equation} \label{eq1}
\begin{split}
e\textsubscript{ni} & = i\textsubscript{na} \sqrt{B} [R\textsubscript{eq}  ||  (1/j\omega C\textsubscript{L})] \\
& =  i\textsubscript{na}\sqrt{B}\frac{R\textsubscript{eq}}{1+j\omega R\textsubscript{eq}C\textsubscript{L}}
\end{split}
\end{equation}
\begin{equation}
\lvert e\textsubscript{ni} \rvert = i\textsubscript{na}\sqrt{B}\frac{R\textsubscript{eq}}{\sqrt{1 + (\omega R\textsubscript{eq}C\textsubscript{L})\textsuperscript{2}}}
\end{equation}
Similar to $e_S$ and $e_L$, the $e_{ni}$ also has LPF like behaviour. In the  temperature range of interest, i.e. \textit{$T_{S}$} $\lesssim$ 1 K, $e_{ni}$ will  be larger than $e_{S}$ and $e_{L}$ over the entire frequency range.\\

(d) Noise contribution ($e_{na}$) due to input voltage noise density: \\
The input voltage noise of the amplifier ($e_{na}$) has two parts, namely, white noise (a frequency independent noise) and the flicker noise (inversely proportional to frequency). The input voltage noise ($e_{na}$) can then be written 
as,\\
\begin{equation} \label{eq1}
\begin{split}
e_{na} = \sqrt{(e_{white}^{2} + e_{flicker}^{2})B} = e_{white} \sqrt{\Big[ 1 +\left( \frac{f_{c}}{f} \right)^{n} \Big]B}
\end{split}
\end{equation}
where, $e_{white}$ is a characteristic parameter of the
amplifier, \textit{$f_{c}$} is the flicker corner frequency and index \textit{n} is in the range 1 to 2
for semiconductor based amplifiers~\cite{flicker}. 
It is evident from Eq. 2.9 that 
there is a strong frequency dependence for \textit{$f<f_c$} due the flicker noise component, while at  higher frequencies ($f>f_c$) $e_{na}$ approaches $e_{white}$. 
Although a $\sqrt T$ dependence for $1/f$ noise is reported in the literature~\cite{flicker_Temp}, in the present study temperature independent value is assumed  since the amplifier is kept at a fixed temperature (T~=~300~K).
\par
Considering all four components as described above,  total  noise at the input ($e_{n-in}$) and output ($e_{n-o}$) of the amplifier with voltage gain $A$ can be written as,
\begin{equation} \label{eq1}
\begin{split}
e_{n-in} = \sqrt{e_{L}^{2} + e_{S}^{2} + e_{ni}^{2} + e_{na}^{2}}
\end{split}
\end{equation}
\begin{equation} \label{eq1}
\begin{split}
e_{n-o} = A \sqrt{e_{L}^{2} + e_{S}^{2} + e_{ni}^{2} + e_{na}^{2}}
\end{split}
\end{equation}

\begin{figure}[h]
\begin{center}
	\includegraphics[scale=0.5]{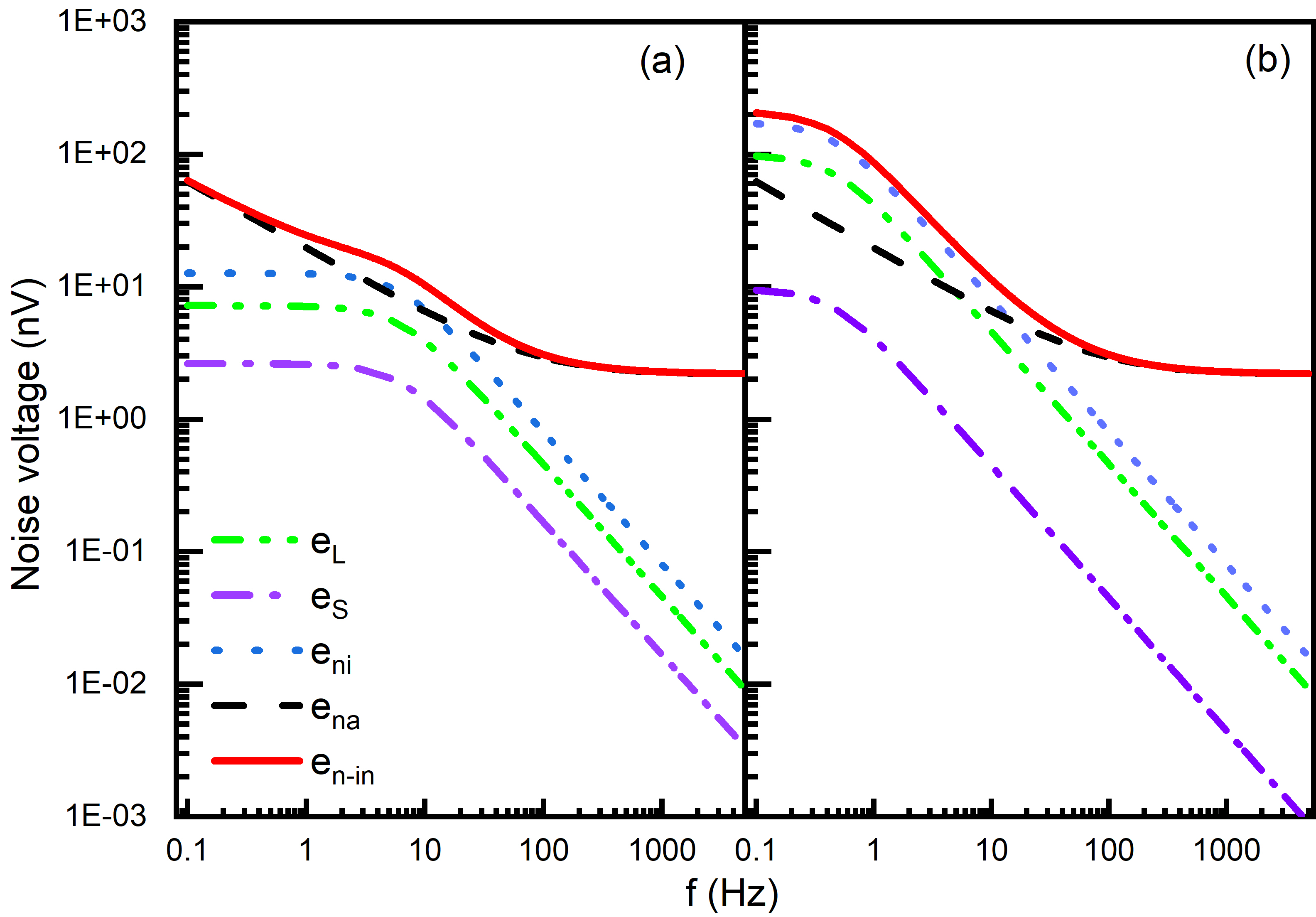}
	\caption{Different noise contributions at the amplifier input for  a)~$R_{S}$ = 25 M$\Omega$  at  50 mK and b)~$R_{S}$ = 350 M$\Omega$ at 50 mK (see text for details)}
	\label{Fig2}
\end{center}
\end{figure}

It should be mentioned that the sensor itself can contribute to the flicker noise, which is not considered here.
Figure~\ref{Fig2} shows calculated individual noise components as a function of frequency for \textit{$R_{S}$}~=~25 and 350~M$\Omega$   at \textit{$T_{S}$} = 50~mK, employing parameters as per the FEMTO datasheet, namely, \textit{$e_{white}$} = 6.9~nV/$\sqrt{\rm Hz}$, \textit{$f_{c}$} = 80~Hz and $i_{na}$ = 1.6~fA/$\sqrt{\rm Hz}$. The index of the flicker noise $n$ is set to be 1  and  $C_L$=1~nF. 
It is evident from the figure that since $T_S < 1~K$,  and  $R_S << R_L$, $e_S$ is smaller than $e_L$.
Further, it can be seen that
the \textit{$e_{S}$, $e_{L}$} and $e_{ni}$ fall off at higher frequency with the same slope, namely, $f_L$, the characteristic frequency of the LPF. 
However,  the  total input noise ($e_{n-in}$) curve can have multiple slopes resulting from the  competition of the input current noise ($f_L/f$)  and the flicker noise (\textit{$f_{c}$/f}) terms.
 In Figure~\ref{Fig2}, the initial \textit{1/f} fall off at very low frequency (\textit{f} < 2 Hz) results from the  flicker noise \textit{$f_{c}$/f}, which dominates the contribution of \textit{$e_{na}$} at low frequencies. While for larger sensor resistance, namely, for $R_S \ge100$~M$\Omega$, the $e_{ni}$ term dominates over $e_{na}$ in the low frequency region  yielding a flat top as  seen in Figure~\ref{Fig2}b.
At higher frequency,  ${f~\gtrsim}$~500~Hz, \textit{$e_{white}$} is the only surviving component of the noise.
Since amplifier characteristic parameters (i.e. \textit{$e_{white}$, $f_{c}$} and \textit{$i_{na}$}) are independent of frequency, variation of total input noise  \textit{$e_{n-in}$} and sensitivity to probe the individual noise components depend on   \textit{$R_{S}$} and \textit{$T_{S}$}. 
For small $R_S$, the $e_{ni}$ contributions is lower than $e_{na}$ at all frequencies (see Figure~\ref{Fig2}a), while for high $R_S$ the $e_{ni}$ component exceeds the $e_{na}$ at low frequencies. However, the value of $R_S$ influences the $f_L$ and therefore the total noise spectrum shows a dominant plateau at $f\le f_L$, characteristic of $e_{ni}$ (e.g. see Figure~\ref{Fig2}b)
It should be pointed out that if  $f_c$ is similar to $f_L$, then it will be difficult to disentangle the flicker noise from the $e_{ni}$ component. 

\section{Noise measurements with NTD Ge sensors and SMD resistors}
\label{chap:Noise measurement with NTD and SMD}
\begin{figure}[h]
\begin{center}
	\includegraphics[scale=0.4]{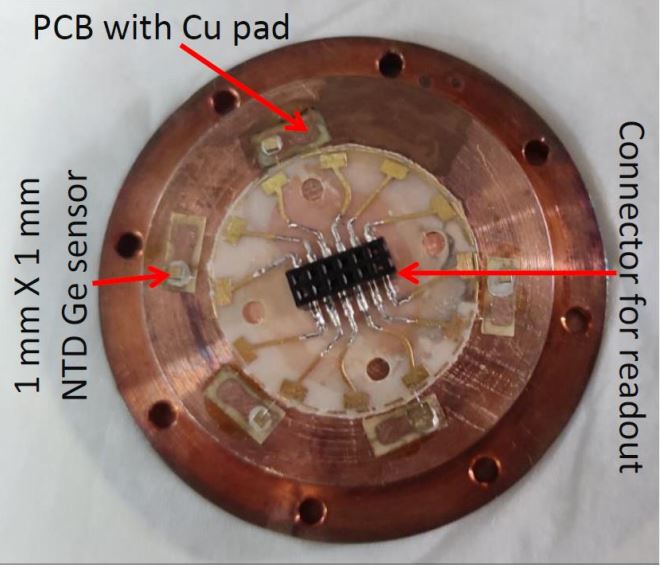}
	\caption{A picture of NTD Ge sensors mounted on copper holder for noise measurement at T~=~20~-~70~mK}
	\label{fig:NTDsetup}
\end{center}
\end{figure}

Noise Measurements  for NTD Ge sensors and SMD resistors were carried out in  separate runs using the CFDR-1200 setup at TIFR. 
Four indigenously fabricated NTD Ge sensors (M713, M715, M716 and M720) were directly 
coupled to the copper holder using GE varnish (see Figure~\ref{fig:NTDsetup}) and the setup  was mounted on the mixing chamber stage of the 
CFDR-1200. All four sensors have similar carrier concentrations and are  identical in dimension (1mm $\times$ 1mm $\times$ 1mm) and with face-type contact~\cite{vatsa_wolte, vatsa_NTD}.  
A set of  SMD resistors, selected from same manufacturing batch with similar packaging, and having resistance  in the range $\sim$ 1 -  100~M$\Omega$ at 300~K, with negative temperature coefficient (i.e. non-metallic behaviour) was chosen for noise measurements as a close match to  NTD Ge for resistance variation with temperature. The SMD resistors were also mounted on a similar copper holder.
The mixing chamber was stabilized at the desired temperature during all the measurements. For simulating the noise in the readout circuit using the model described in the previous section, knowledge of  $R_S$ and $C_L$ is an essential input. Measurements of $R_S$, $C_L$ and noise are described in the following.  \\

\subsection{$R_{S}$ and $C_{L}$ measurement}
\label{$R_{S}$ and $C_{L}$ measurement}

Sensor resistance measurements were done either with AVS-47B~\cite{avs} resistance bridge (for $R_S$ < 2~M$\Omega$, typically $T_S > 100$~mK) or with National Instrument-based Data Acquisition (NI PXI-6281 DAQ) system with FEMTO amplifier (for $R_S$ > 2~M$\Omega$, typically $T_S < 100~$mK). In case of measurements with NI-DAQ, 
the sensor $R_S$ was connected between the terminals 'a' and 'b' as shown in Figure~\ref{fig:ckt} and a square wave excitation was applied to the sensor through a high bias resistor (\textit{$R_L$}=20 G$\Omega$) connected in series with the sensor. The amplitude of the input voltage was optimally chosen to get an acceptable signal to noise ratio
(SNR), while minimizing the self heating of the sensor.
The  voltage response at output of the amplifier with gain $A$~=~60~dB, was recorded for a chain of 60 input voltage pulses. From the histogram of the output peak-to-peak voltage $V_o$ recorded for all pulses, mean amplitude $<V_o>$ was extracted and the sensor resistance was then calculated as 
    \begin{equation}
 	    R_S~=~\frac{V_S}{V_{in} - V_S} \times R_L
 	\end{equation} 
where $V_{in}$ is the  voltage drop across $R_S + R_L$ and $V_S =<V_o>/A$ is the voltage drop across the $R_S$.

\begin{figure}[h]
\begin{center}
	\includegraphics[scale=0.55]{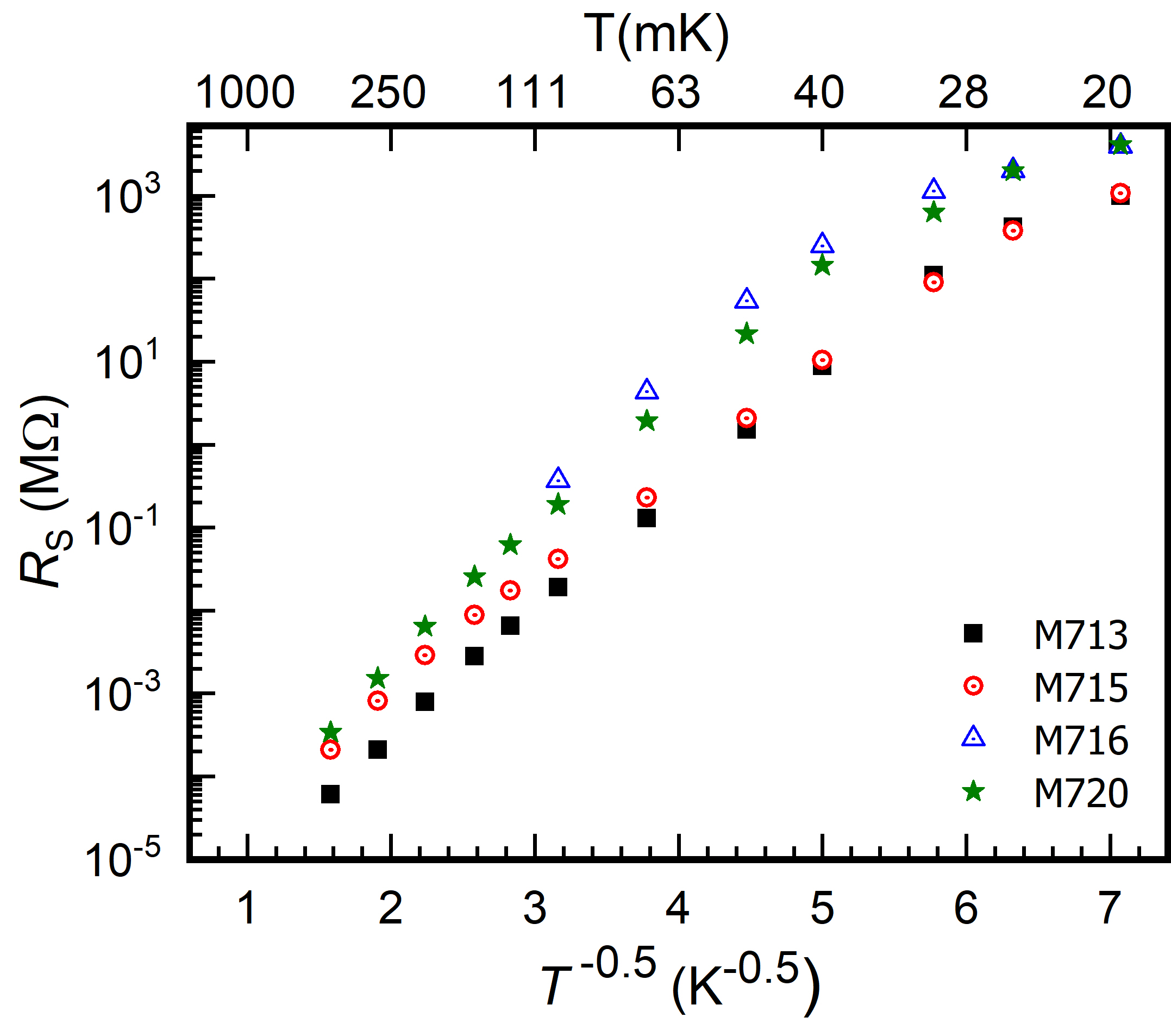}
	\caption{Variation of $R_{NTD}$ as a function of T showing Mott behaviour.}
	\label{fig:Mott}
\end{center}
\end{figure}

For measurements with AVS-47B, the excitation voltage was fixed at a 30~$\mu$V and the data was averaged over 25 minutes. 
It can be seen from Figure~\ref{fig:Mott} that resistances of all four NTD Ge sensors follow the expected Mott-like behaviour $ R = R_{0} ~exp\left(\sqrt{\frac{T_{0}}{T}}\right)$~\cite{mott}  with <$T_{0}$> = 10.13 $\pm$ 0.02 K over a temperature range of 20 - 400 mK.

\begin{figure}[h]
\begin{center}
	\includegraphics[scale=0.5]{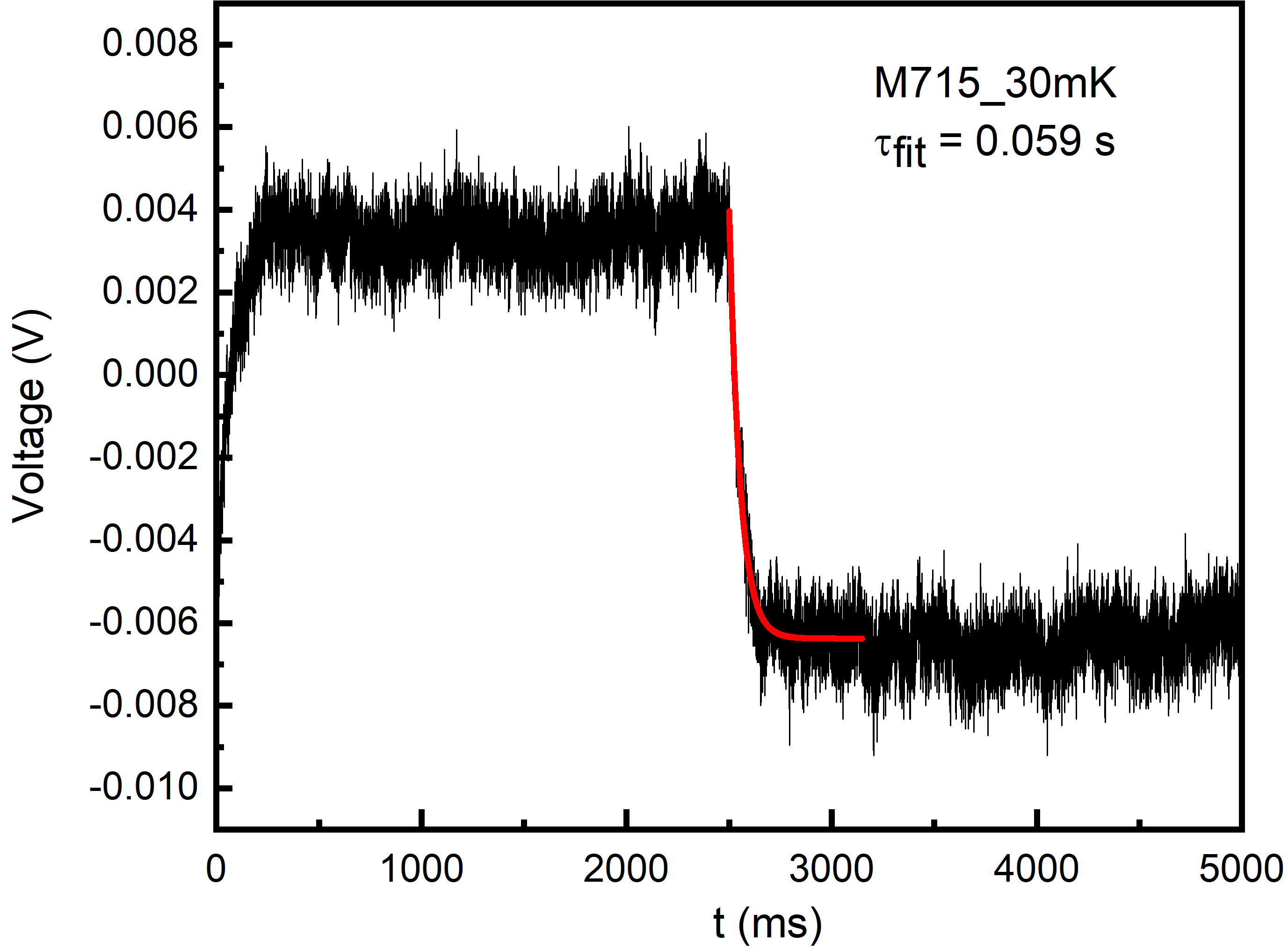}
	\caption{A typical example of an output waveform across the sensor resistor $R_S$~=~87~M$\Omega$. Exponential decay fit to the falling edge giving $\tau_{fit}$~=~0.059 s corresponding to $C_{L}$~=~0.7~nF is also shown.}
	\label{expoDec}
\end{center}
\end{figure}

\begin{figure}[h]
\begin{center}
	\includegraphics[scale=0.5]{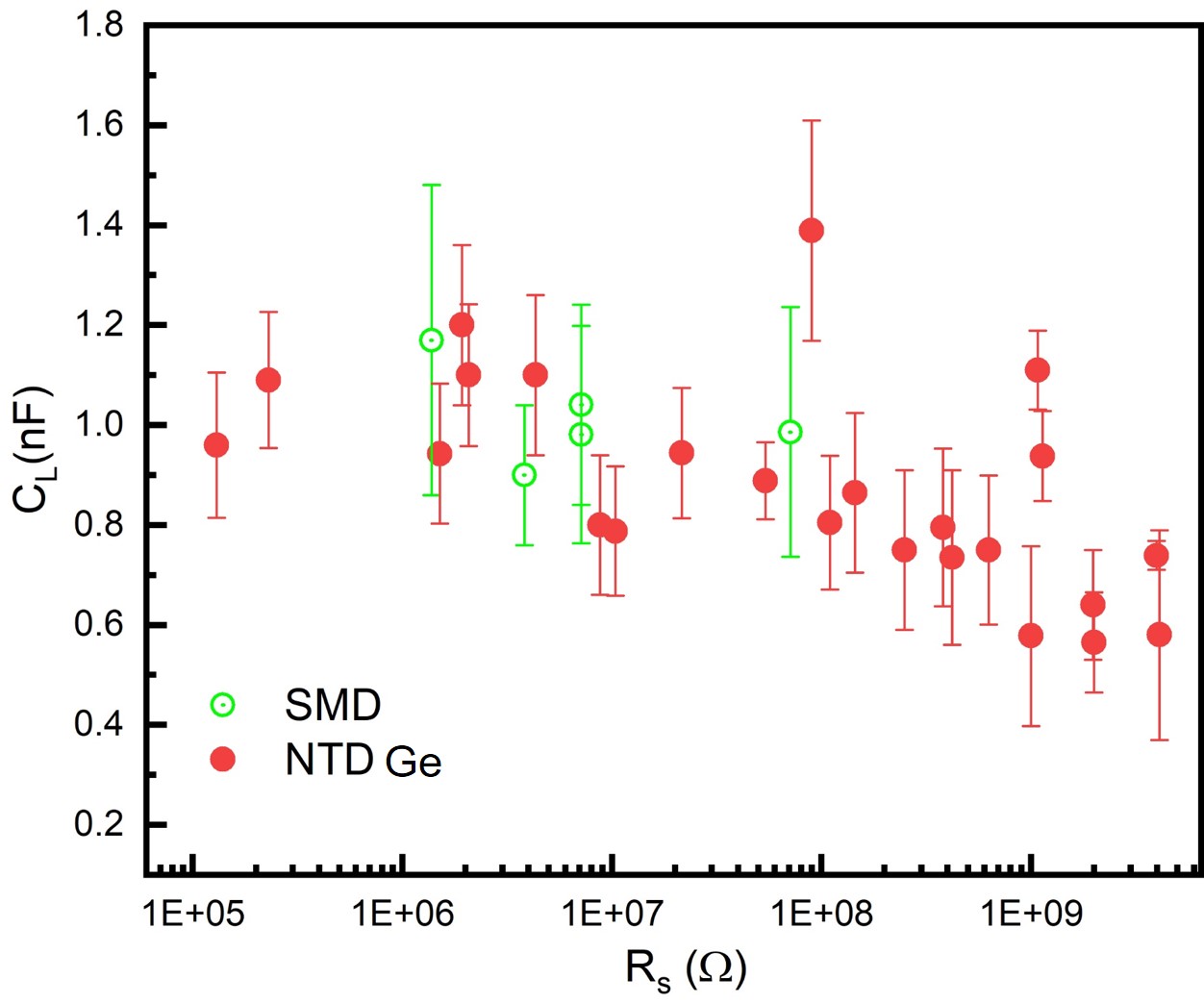}
	\caption{Distribution of estimated capacitance values for various $R_S$ (SMD and NTD Ge sensors).}
	\label{varCap}
\end{center}
\end{figure}

The $C_{L}$ was estimated by fitting the decay time ($\tau$, i.e. RC time constant) of the falling edge of individual output pulses at the output of the amplifier (corresponding to the square wave input).  As mentioned earlier, the measurements were done using very low bias voltage to avoid any self heating and hence the resistance was assumed to be a constant during the rise and fall of the square pulse excitation.
A typical example of an exponential decay fit is shown in Figure~\ref{expoDec}.  For each $R_{S}$ at given $T_{S}$, average $\tau$ was obtained by fitting  12 pulses  and the standard deviation ($\sigma$) is taken as the uncertainty. The mean capacitance   \textit{$C_{L}$} = $\tau$$_{mean}$/\textit{$R_S$}, thus obtained are shown in Figure~\ref{varCap} for various \textit{$R_{S}$}. For further analysis, $C_{L}$ is taken to be  0.85~nF, which is  the weighted mean for the NTD Ge sensor data. It was verified that simulated noise is not very sensitive to $C_{L}$ and upto 20\% variation in $C_{L}$ did not produce any observable change in e$_{n-o}$.


\subsection{Noise measurement}
\label{Noise measurement}
\begin{figure}[h]
\begin{center}
	\includegraphics[scale=0.53]{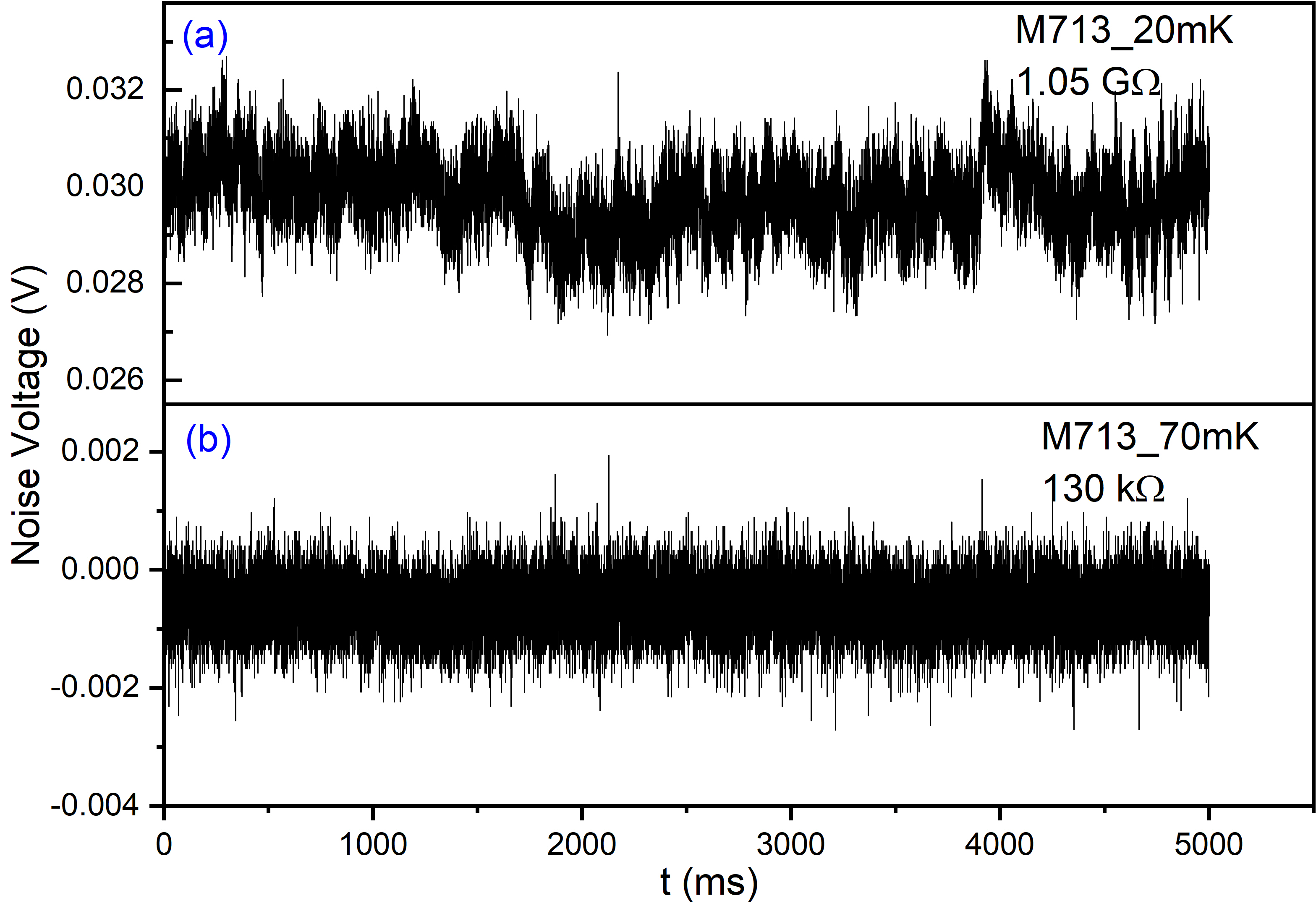}
	\caption{Typical noise data for NTD Ge sensor M713 at $T$~=~20~mK (top panel) and $T$~=~70~mK (bottom panel)}
	\label{noise}
\end{center}
\end{figure}

Noise measurements were done with NI DAQ over a temperature range of 20~-~70~mK for NTD Ge sensors, spanning a range of $R_S$ over three orders of magnitude. A set of SMD resistors was selected to cover a similar range in $R_S$ over a higher temperature range of $\sim 3~-~13$~K. Ideally, noise measurements are done with zero bias voltage. However, a small DC voltage was applied to compensate any residual offset at output of the sensor arising due to thermoEMF~\cite{thermo} and other effects. A total of 100~k samples were acquired in a time window of 10~s using a sampling rate of 10~kHz,  with bandwidth of the amplifier kept at 1~kHz. 
A typical 5 sec noise data for NTD Ge sensor M713 at T = 20~mK and at T = 70~mK are shown in Figure~\ref{noise}. For each measurement (i.e. for given R$_S$, T$_S$), data were acquired in 12 time windows of 10 sec each (total length 120 s). The data in 12 time windows were averaged over to construct a single 10~s time domain data. It should be mentioned that at lower temperatures, drifts and wobbles in the baseline become visible, which correspond to ultra low frequency components and can be a limiting factor. Averaging over multiple time windows helps to improve the statistics.

\section{Optimization of model} 
\label{Analysis and simulation}
\begin{figure}[h]
\begin{center}
	\includegraphics[scale=0.5]{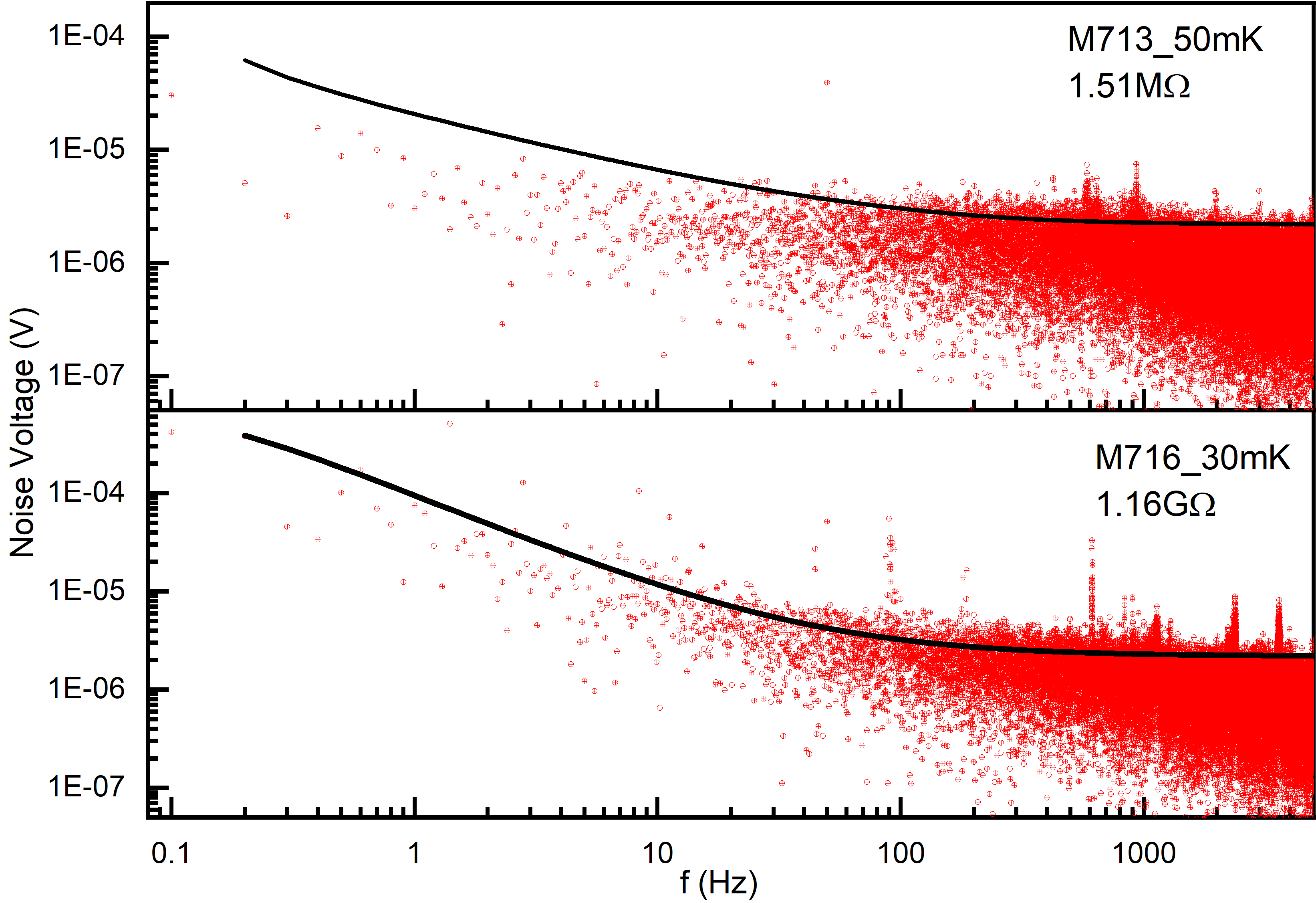}
	\caption{A comparison of simulated noise with parameters taken from the datasheet (solid black line) with measured noise data (red dots) for NTD Ge sensors. It is evident that \textit{$f_{C}$ }~=~80~Hz over-predicts the noise at $f \lesssim$ 100~Hz.}
	\label{fig8}
\end{center}
\end{figure}

For further analysis and comparison with simulated noise spectra, 12 independently acquired 10~s long noise data were averaged to a single time window, 
 to improve the statistics and thus the quality of the acquired data (as explained in the previous section). A discrete Fast Fourier Transform (FFT) of noise data was generated for the averaged  10 sec time data (as mentioned above) using a standard FFT algorithm and further analysis was carried out in frequency domain. The resolution bandwidth $(B)$ for the FFT was set at 0.1~Hz and correspondingly time window was chosen to be 10~s. The simulated noise data was folded with a low pass filter function to take into the account the bandwidth of the amplifier (1~kHz). 
For a given \textit{$R_{S}$} and $T_{S}$, the noise essentially depends on  three parameters --- \textit{$i_{na}$, $f_{c}$ and $e_{white}$}.
Initial noise simulations were done with datasheet values for these  parameters and  two examples for different $T_S$ and $R_S\sim$ few M$\Omega$ are shown in Figure~\ref{fig8}.
From the figure, it is evident that in low frequency window (f < 10~Hz), the model over-predicts the experimental data. Additionally, for \textit{$f_{c}$} = 80~Hz (as per the  datasheet) and \textit{n}~=~1, the  \textit{$e_{na}$} will be the dominant contribution in the entire frequency region for   \textit{$R_{S}$} $\leq$ 100~M$\Omega$, and  is expected to show  \textit{1/f} behaviour till $f_c$. This behaviour is not observed in noise data at 40~-~50 mK, which shows a rather flat trend at low frequency (e.g. see Figure~\ref{fig8}). In most of the data  at $T\sim$~50~-~70~mK, where \textit{$R_{S}$}~$\le$~10~M$\Omega$, the initial \textit{1/f} component exhibits  much smaller slope as compared to that expected for \textit{$f_{c}$} = 80~Hz. 
Hence, an optimization of model parameters is needed for a better description of the data over a range of $R_S-T_S$. It should be mentioned that a few discrete  peaks are also observed  in the measured spectra at  harmonics of 50 Hz, high frequency due to vacuum pumps \& instrumentation, and in some cases harmonics of 1.4~Hz (frequency of the pulse tube cooler). However, these sharp noise peaks occurring at very specific frequencies do not influence the overall analysis of the underlying noise. \par

As seen  from  Figure~\ref{Fig2}, the parameter \textit{$e_{white}$} corresponds to the saturation noise value at higher frequencies and  can be unambiguously  extracted  irrespective of $R_{S}$ and $T_{S}$. 
For extracting other two parameters (i.e. \textit{$f_c$} and \textit{$i_{na}$}), the $R_S$ were carefully chosen to achieve maximum sensitivity for reliable fit, as explained earlier in the section 2.
A total of 8 datatsets of NTD Ge sensors with $R_S\sim$~1~-~50~M$\Omega$ and  $T_S\sim$~40~-~70 mK were considered. 
The simulated noise was fitted to the experimental data in the frequency window 0.2 to 5000 Hz,  after incorporating the amplifer response.
The data at points $f$~=~0 and 0.1 Hz were intentionally not included in the fitting region, since the first two points in the FFT can be influenced by any residual DC offset in the time spectra. 
The fitting to simultaneously extract  \textit{$i_{na}$, $f_{c}$ and $e_{white}$} was done using Root analysis framework~\cite{root}. \\ 

\begin{figure}[h]
\begin{center}
	\includegraphics[scale=0.55]{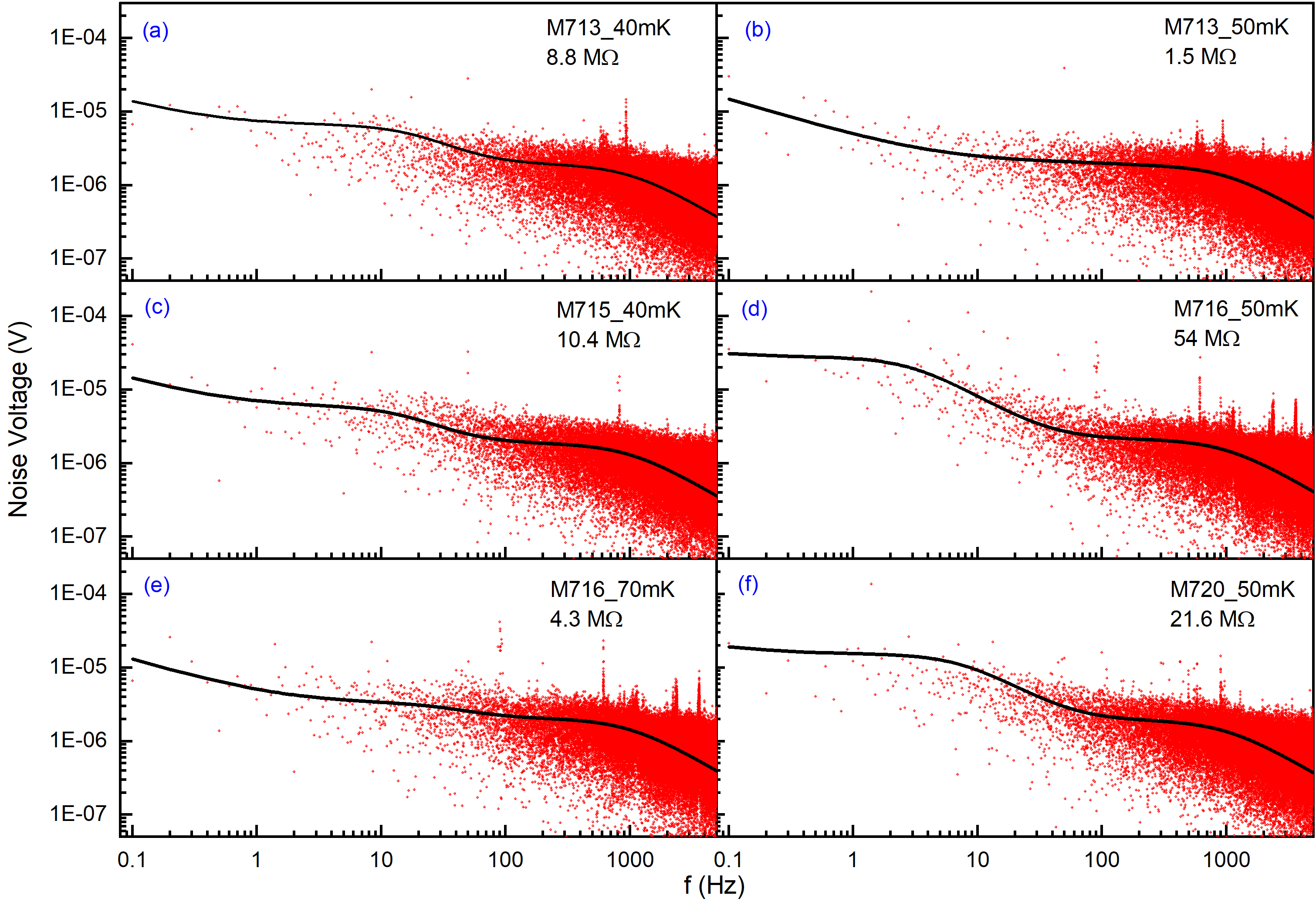}
	\caption{Simulated best fit noise curves (solid black line) with experimental data (red dots) for different NTD Ge sensors.  Fit parameters are listed in Table~\ref{tab:Table3} (see text for details).}
	\label{fig:fitcomp}
\end{center}
\end{figure}
Figure~\ref{fig:fitcomp} shows some of the data together with the best fit simulations.
As mentioned earlier, the multiple slopes visible in simulated noise spectra result from competing amplitudes and slopes of \textit{$f_c$/f} and \textit{$f_L$/f}  terms.
For the datasets with $R_S\sim$  1-10~M$\Omega$ at \textit{$T_S$}$\sim$~50~-~70~mK, contributions of  $e_S$  and $e_L$ are relatively small as compared to the other terms. The \textit{$e_{na}$} term dominates over other contributions in the lower frequency window ($f$ < 2~Hz) and hence  is strongly sensitive to $f_c$. 
 As $e_{na}$ term decreases rapidly with increasing $f$,  $e_{ni}$ becomes a dominant contributor and leads to a flat noise in the 2-20~Hz window. 
  The observed fall beyond this frequency is due to the \textit{$f_L$/f} term. At $f~>1~$kHz, the amplifier bandwidth response results in the observed decreasing trend in the spectrum. It may be pointed out that in Figure~\ref{fig:fitcomp}(b, e), the \textit{$f_L$/f} slope is not visible because of very small $R_S$.

\begin{table}[h!]
  \begin{center}
    \caption{Model  parameters obtained from fitting the NTD Ge sensor data}
    \smallskip
    \label{tab:Table3}
    \begin{tabular}{|c|c|c|c|c|c|}
	\hline
      {Sensor Tag} & {$R_S$} & {$f_{c}$} & {$e_{white}$} &{$i_{na}$} & {$\chi$${^2}$$_{red.}$} \\
       & M$\Omega$ & Hz & nV/$\sqrt{Hz}$ & fA/$\sqrt{Hz}$ &   \\
      \hline
      M713\_40mK & 8.8 & 4.2 $\pm$ 0.44 & 5.99 $\pm$ 0.021 & 1.95 $\pm$ 0.018 &  0.12\\
      M713\_50mK & 1.5 & 3.9 $\pm$ 0.19 & 5.65 $\pm$ 0.018 & 2.15 $\pm$ 0.075 & 0.12\\  
      M715\_40mK & 10.4 & 5.3 $\pm$ 0.27 & 5.77 $\pm$ 0.018 & 1.28 $\pm$ 0.022 &  0.12\\
      M715\_50mK & 2.1 & 4.4 $\pm$ 0.25 & 5.67 $\pm$ 0.020 & 1.04 $\pm$ 0.010 & 0.14\\
      M716\_50mK & 54.0 & 4.5 $\pm$ 0.70 & 6.61 $\pm$ 0.230 & 1.30 $\pm$ 0.190 & 0.55\\
      M716\_70mK & 4.3 & 3.9 $\pm$ 0.28 & 6.36 $\pm$ 0.023 & 1.25 $\pm$ 0.067 & 0.18\\
      M720\_50mK & 21.6 & 3.9 $\pm$ 0.98 & 5.99 $\pm$ 0.032 & 1.97 $\pm$ 0.017 & 0.25\\ 
      M720\_70mK & 1.9 & 4.5 $\pm$ 0.22 & 5.7 $\pm$ 0.018 & 2.15 $\pm$ 0.075 & 0.11\\
      \hline
      Mean & & 4.3 $\pm$ 0.4 & 5.97 $\pm$ 0.05 & 1.63 $\pm$ 0.06 &  \\
	\hline
    \end{tabular}
  \end{center}
\end{table}

Table~\ref{tab:Table3} lists the extracted fit parameters with fitting errors and $\chi^2_{red}$.
The $\chi^2_{red}$ is computed as,
\begin{equation}
    \chi^2 _{red.} = \frac{1}{N_{DF}} \sum_{i} \frac{1}{\sigma _i ^2}[(e_{sim})_i  - (e_{data})_i]^2
\end{equation}
where $N_{DF}$ is number of degrees of freedom (= 49997), $e_{sim}$ and \textit{$e_{data}$} refer to simulated and measured noise voltage, respectively, in the $i^{th}$ frequency bin. 
 The $\sigma_i$ is taken as $e_{white}$ (value obtained from the fit), which reflects the noise fluctuation in each bin. \par  

The mean values for $f_c$, $f_{white}$ and $i_{na}$ listed at the bottom of the table are taken as optimal values for further simulations. It is important to note that $e_{white}$ and $i_{na}$ are similar to datasheet values, but the $f_c$ value is drastically lower. The reason for this discrepancy is not understood. It should be mentioned that independent measurements of FEMTO DLPVA-100F were also performed with low value resistors (1, 10 and 100 $\Omega$)  connected at the input, which also yielded $f_c$ $\sim$ 4~Hz. 

\begin{figure}[h]
\begin{center}
	\includegraphics[scale=0.55]{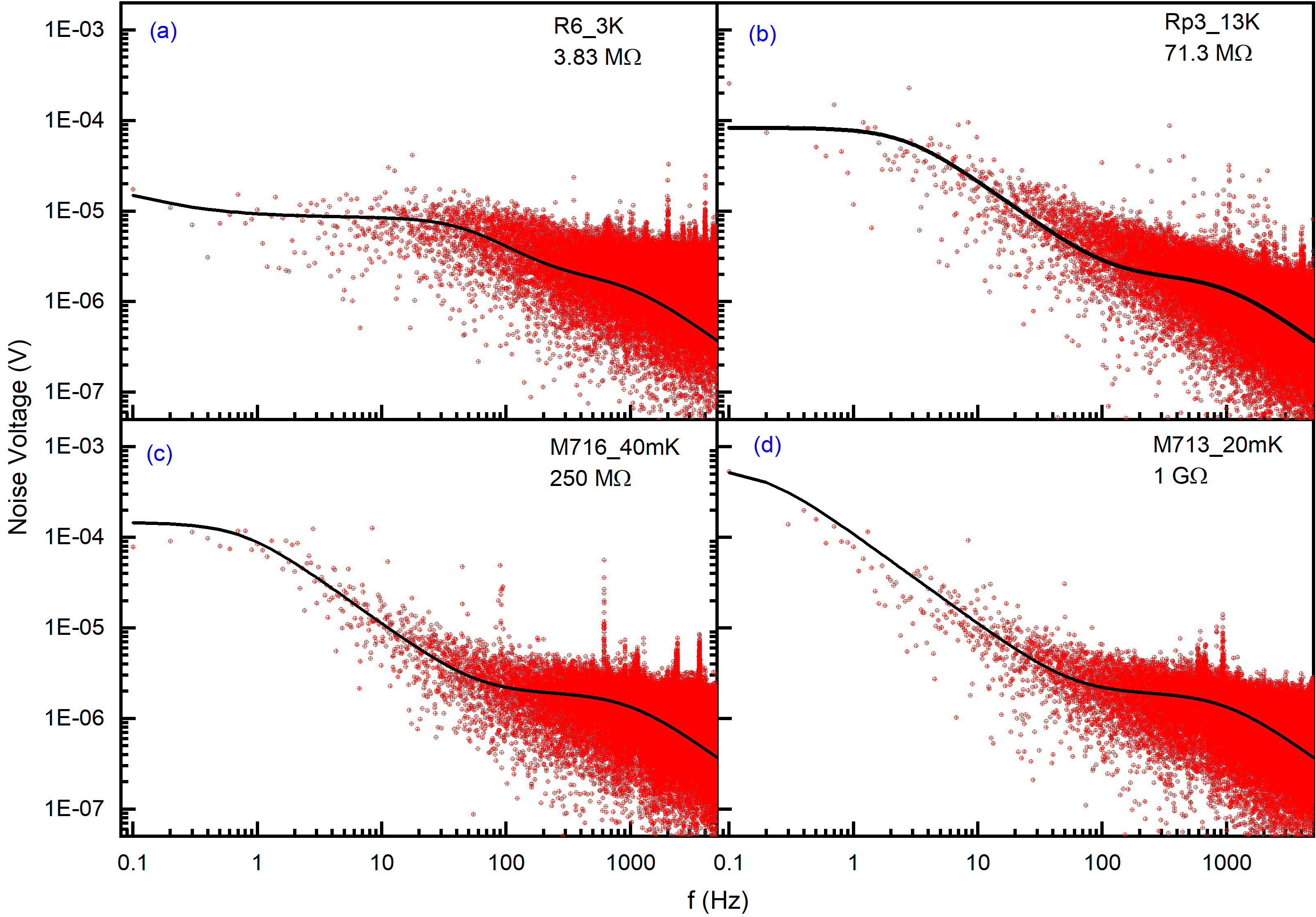}
	\caption{A comparison of simulated noise using optimized parameters (\textit{$f_c$}~=~4.3Hz, \textit{$i_{na}$}~=~1.63fA/$\sqrt{Hz}$ and \textit{$e_{white}$} = 5.97nV/$\sqrt{Hz}$) and experimental data for various sensors in different $R_S$~-~$T_S$ range. Overall good agreement is visible (see text for details).}
	\label{Fig10}
\end{center}
\end{figure}

To verify the model, the simulated noise spectra  with these optimized parameters were compared with various NTD Ge sensor and SMD resistor datasets at different temperatures. A few examples  of simulated noise together with data  for $R_S\sim$~10~-~1000~M$\Omega$ are 
shown in Figure~\ref{Fig10}. 

The good agreement between data and simulations at a higher $T_S$ value (3~K) is clearly visible in Figure~\ref{Fig10}a. The flat noise at lower frequency and subsequent fall in the data for  \textit{$R_S$}~=~71.3 and 250 M$\Omega$ (panel b and c of Figure~\ref{Fig10}) are well reproduced by the simulation. Similarly, for high $R_S$ value (Figure~\ref{Fig10}(d), 1~G$\Omega$) the observed fall in data is duly described by the dominant noise term $e_{ni}$ of the model.

Another figure of merit for comparison of  the data and simulation is the RMS value of noise voltage. From the model and data FFT spectrum, the RMS noise voltage $e_{rms}$  is computed as  
\begin{center}
\begin{equation} \label{eq1}
\begin{split}
e_{rms} = \sqrt{\int_{f_{l}}^{f_{h}}\frac {e_{i}^{2}}{B} df}
& = \sqrt{\sum_{i} e_{i}^{2}}
\end{split}
\end{equation}
\end{center}
where (\textit{$f_{h}$ - $f_{l}$}) is the acquisition bandwidth of noise and \textit{df} is the resolution bandwidth (which is same as B in the present case).   Simulated and measured $e_{rms}$ for frequency window 0.2~-~1000~Hz are given in the Table~\ref{tab:Table2} for NTD Ge sensors spanning temperature range 20 to 70 mK. Additionally, to discount the effect of the DC offset and low frequency wobble in the experimental data, $e_{rms}$  only in the 5.1~-~1000~Hz  window is also shown.
Table~\ref{table3SMD} shows data for SMD resistors in the temperature range 3 to 13~K. 
It should be mentioned that the flicker noise contribution in the SMD resistors is expected to be different from that in  NTD Ge sensors, given the difference in the volume and packaging. However, for the range of $R~-~T$ under consideration, $e_{ni}$ and $e_S$ are the dominant terms and measured data is not sensitive to $f_c$.
 From both the tables it is evident that overall agreement between model and data (i.e. $<~e_{rms}^{data}$/ $e_{rms}^{sim}>$) is within 15$\%$ for the frequency range of 5.1-1000~Hz. Thus, it can be concluded that the analytical noise model with optimized parameters  is able to describe data over a wide range of $R_S$ and $T_S$. 
 
\begin{table}[h!]
  \begin{center}
    \caption{A comparison of simulated noise,  $e_{rms}^{sim}$ (with optimized parameters from Table~\ref{tab:Table3}) and measured data $e_{rms}^{data}$ for NTD Ge sensors.}
    \smallskip
    \label{tab:Table2}
    \begin{tabular}{|c|c|c|c|c|c|}
	\hline
      {sensor tag} & $R_S$  & $e_{rms}^{sim}$ (mV) & $e_{rms}^{data}$ (mV) &$e_{rms}^{sim}$ (mV) & $e_{rms}^{data}$(mV) \\
       & M$\Omega$ & (0.2-1000)Hz & (0.2-1000)Hz & (5.1-1000)Hz &(5.1-1000)Hz \\
      \hline
      M713\_20mK & 1050 & 0.78 & 0.5 & 0.23 & 0.27\\
      M713\_30mK & 111 & 0.37 & 0.35 & 0.23 & 0.23\\
      M713\_70mK & 0.13 & 0.17 & 0.17 & 0.17 & 0.16\\
      M715\_20mK & 1140 & 0.79 & 0.52 & 0.23 & 0.26\\
      M715\_30mK & 90.4 & 0.34 & 0.34 & 0.23 & 0.28\\
      M715\_70mK & 0.23 & 0.17 & 0.18 & 0.17 & 0.18\\
      M716\_20mK & 4010 & 0.89 & 0.89 & 0.23 & 0.31\\ 
      M716\_30mK & 1160 & 0.79 & 0.76 & 0.24 & 0.29\\
      M716\_40mK & 250 & 0.5 & 0.67 & 0.23 & 0.31\\
      M720\_20mK & 4100 & 0.88 & 2.07 & 0.24 & 0.28\\
      M720\_30mK & 650 & 0.69 & 0.95 & 0.23 & 0.26\\
      M720\_40mK & 145 & 0.41 & 0.69 & 0.22 & 0.24\\
	\hline
    \end{tabular}
  \end{center}
\end{table}

\begin{table}[h!]
  \begin{center}
    \caption{A comparison of simulated noise,  $e_{rms}^{sim}$ (with optimized parameters from Table~\ref{tab:Table3}) and measured data $e_{rms}^{FFT}$ for SMD resistors.}
    \smallskip
    \label{table3SMD}
    \begin{tabular}{|c|c|c|c|c|c|}
	\hline
      {sensor tag} & $R_S$  & $e_{rms}^{sim}$ (mV) & $e_{rms}^{data}$ (mV) &$e_{rms}^{sim}$ (mV) & $e_{rms}^{data}$(mV) \\
       & M$\Omega$ & (0.2-1000)Hz & (0.2-1000)Hz & (5.1-1000)Hz &(5.1-1000)Hz \\
      \hline
      R6\_3K & 3.83 & 0.28 & 0.36 & 0.27 & 0.36\\
      Rp1\_13K & 1.38 & 0.46 & 0.44 & 0.46 & 0.44\\
      Rp2\_13K & 7.15 & 0.49 & 0.54 & 0.47 & 0.52\\
      Rp3\_13K & 71.3 & 0.55 & 0.60 & 0.34 & 0.43\\
      Rp6\_13K & 7.15 & 0.49 & 0.49 & 0.46 & 0.47\\
      
      \hline
    \end{tabular}
  \end{center}
\end{table}

\begin{figure}[h]
\begin{center}
	\includegraphics[scale=0.55]{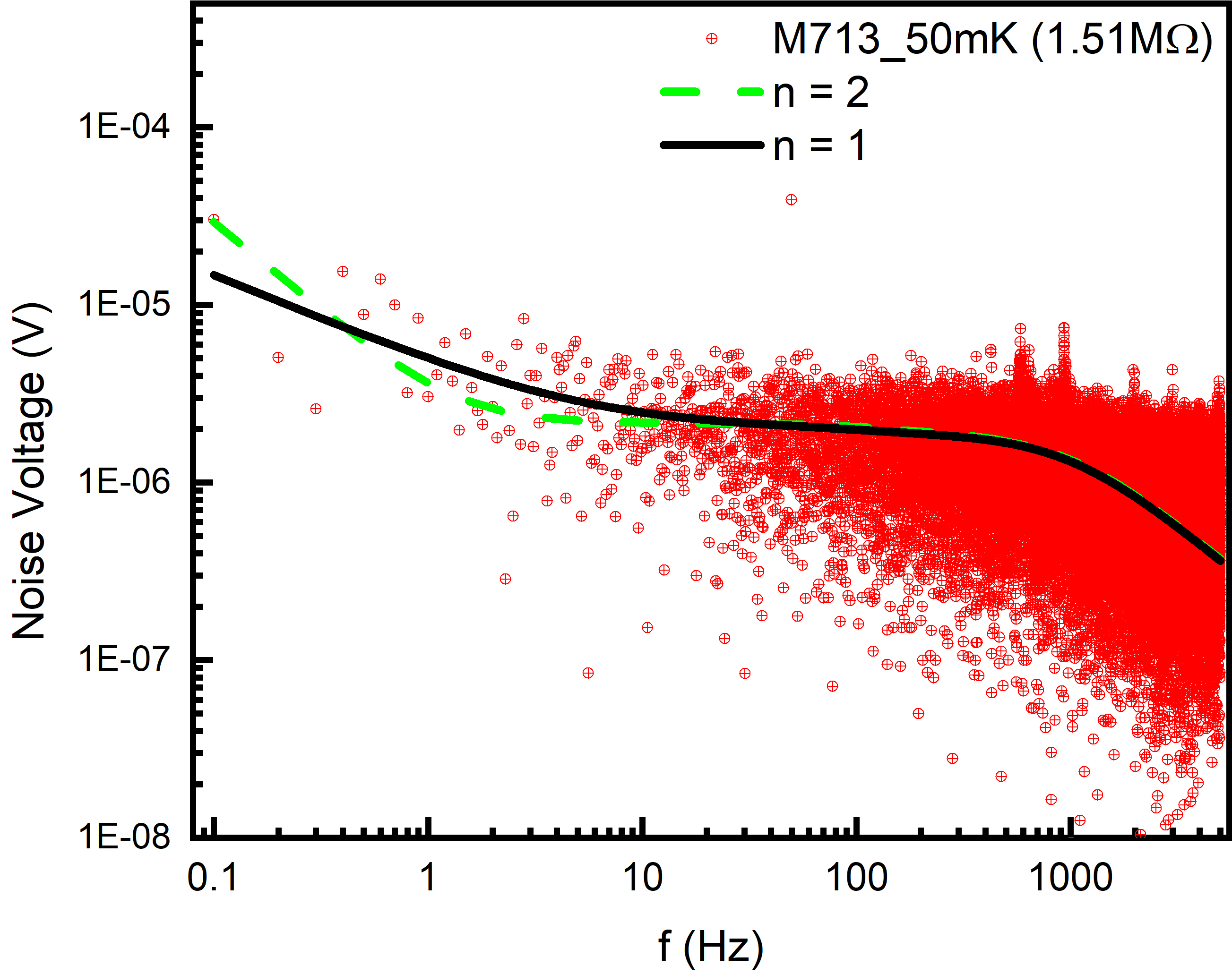}
	\caption{A comparison of simulated noise for flicker noise index $n~=$~1 and 2  for $R_S$~=  1.51~M$\Omega$, $T_S$~= 50~mK.}
	\label{fig:ncomp}
\end{center}
\end{figure}

As mentioned in section~\ref{section2}, the index $n$ of the flicker frequency can take values between 1 to 2 and $n$~=~1, which corresponds to the slowest fall, was used.
It should be pointed out that index $n$~= 1 and 2 effectively encompasses flicker and shot noise. To assess the impact of $n$ on noise parameters, fitting was also done with $n$=2 in Eq.~2.9. It is observed that the parameters $e_{white}$ and $i_{na}$ are not influenced by choice of $n$, but $f_c$ varies as expected. 
The  simulated noise spectra with $n$=1 and 2 for best fit parameters are shown in Figure~\ref{fig:ncomp}.  It can be seen that differences are visible only at very low frequency region ($f<10$ Hz) and the $e_{rms}$ values are nearly identical 0.173 and 0.177 mV  for  $n$=1 and 2, respectively. The present data cannot distinguish between $n$=1 or 2.

Except \textit{$e_{na}$}  other noise factors have  \textit{R$_S$} dependence, with \textit{$e_{ni}$} as a leading term proportional to the $R_S$. A high value of $R_S$ is essential for good sensitivity in the bolometer signal. However, readout circuit noise and noise due to other factors like harmonics of pulse tube cooler, vacuum pumps and instrumentation, etc. also increase with increasing sensor resistance. Thus, an optimal value of $R_S$ is desirable to enhance the signal, while keeping overall noise within acceptable window. 
Employing the present model, it is seen that  for  $R_S\lesssim$ 100~M$\Omega$, \textit{$e_{na}$} is the  dominant contribution at all frequencies.
While for $R_S~\gtrsim~100~$M$\Omega$ both \textit{$e_{ni}$} and \textit{$e_{L}$} will be dominant factors.
However, for $R_S$ $\sim$~100~-~350~M$\Omega$ the \textit{$e_{ni}$} and \textit{$e_{na}$} contributions will be of the similar magnitude at low frequency. 
For $R_{NTD}\sim$ 1~G$\Omega$, the expected noise  is shown in 
Figure~12, which gives $e_{rms}^{sim}\sim$~0.5~mV  at the amplifier output.
Hence, from noise point of view it is suggested that $R_{NTD}$ should be smaller than 1~G$\Omega$. 
Since the dependence of the resistance of NTD Ge sensor on temperature is described by Mott behaviour (see section 3.1), the sensitivity of NTD Ge sensor~\cite{sensitivity} i.e. $\frac{dR}{dT}$ is proportional to  $R_S \sqrt{\frac{T_0}{T}}$. 
Therefore, for better sensitivity, large $R_s$ is desirable.
Thus, $R_{NTD}$ in the range of 0.5~-~1~G$\Omega$ will be preferable to achieve a good sensitivity and low noise. 

\begin{figure}[h]
\begin{center}
	\includegraphics[scale=0.45]{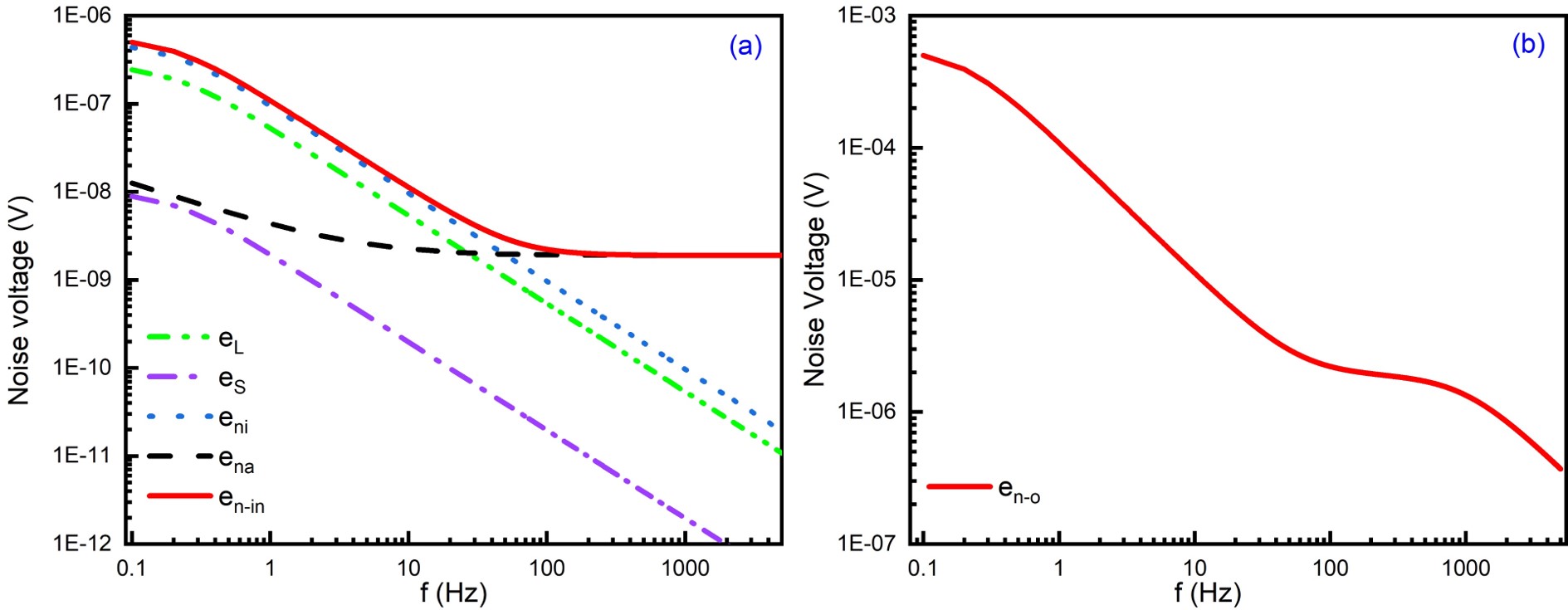}
 	\caption{Simulated noise spectrum for $R_{NTD}$~=~1~G$\Omega$ at 20 mK with optimized parameters: (a)~individual contributions and total noise at the amplifier input, (b)~total noise at the amplifier output (inclusive of the amplifier response)}
	\label{fig:RNTD1G}
\end{center}
\end{figure}

\section{Conclusion}
\label{Conclusion}
An analytical model for  bolometer readout circuit has been presented to understand the effect of noise from different circuit components. The parameters of  the model (\textit{$e_{white}$}, \textit{$i_{na}$} and \textit{$f_c$}) have been optimized to  fit the observed data of NTD Ge sensors. 
While empirical values of the  parameters \textit{$e_{white}$}, 
\textit{$i_{na}$}  are consistent with the 
datasheet values, the flicker corner frequency \textit{$f_{c}$} is observed to be significantly smaller (4.3~Hz) than the specified value of 80~Hz. 
The optimized  model has been tested for  NTD Ge sensors over a wide range of $R_S~=~$130~k$\Omega$~-~4.1~G$\Omega$ in 20~-~70~mK and for SMD resistors in the range $T~=~3~$-$~13~$K,  
and is found to describe FFT of the measured data very well.
The model can be used to  predict the value of voltage noise \textit{$e_{rms}$}. The model is general enough and can be extended to any boloemeter readout circuit or amplifier. 
From noise considerations, the desired range for resistance of NTD Ge sensor is shown to be 0.5~-~1~G$\Omega$.
Further, it is shown that the current noise from amplifier is the  leading source of the noise for the relevant range of $R_S\sim$~350~-~1000 M$\Omega$ for NTD Ge sensor below 100~mK. Consequently, A suitable low temperature amplifier ($\sim$~100~-~120~K) with small $i_{na}$ and low $f_c$ is desirable for the cryogenic bolometer readout.

\acknowledgments
We thank Mr. K. V. Divekar for assistance during measurements. This work is supported by the Department of Atomic Energy, Government of India, under Project No. RTI4002.

\end{document}